\documentclass[11pt]{article}
\usepackage{preprint}

\runningtitle{Mumcu et al.}

\ptitle{Submission Responsibility Matters: \\ Role-Aware Submission Quotas under Coauthorship}

\pauthor{Furkan Mumcu}{1}{furkan@usf.edu}
\pauthor{Yasin Yilmaz}{1}{}
\paffiliations{$^{1}$Department of Electrical Engineering, University of South Florida}

\pabstract{%
Author-level submission quotas are increasingly used to control growing peer-review load. Recent coauthorship-sensitive quota rules improve over fixed per-author limits by reducing the quota cost of multi-author submissions, often using harmonic authorship-credit models to prevent simple author-list padding. However, these rules conflate three distinct quantities: review burden, authorship credit, and submission responsibility. As a result, they can penalize genuine solo-authored work, treat all coauthors as equally responsible for a submission, and create bottlenecks for student-led papers when a faculty advisor appears on multiple unrelated submissions.

We argue that submission quotas should be designed around the responsibility structure of a paper rather than only its number of coauthors. We formalize desiderata for quota rules, including venue-load control, padding resistance, role sensitivity, solo neutrality, and student non-blocking. We then propose a role-aware quota framework that assigns author-specific quota costs based on constrained roles such as lead author, regular coauthor, and designated advisor. The framework includes fixed, per-capita, and harmonic-style rules as special or limiting cases, while allowing venues to distinguish lead authors, corresponding authors, advisors, and peripheral contributors. We show how simple role constraints can preserve resistance to manipulation while avoiding several structural disadvantages of coauthor-symmetric quota rules. Our analysis suggests that role-aware quota mechanisms provide a more faithful and flexible foundation for managing peer-review load under modern collaborative authorship.
}

\begin{document}

\makepreprinttitle

\section{Introduction}

Peer-review systems are under increasing pressure from growing submission volume and reviewer scarcity~\cite{publons2018global,adam2025peerreviewcrisis,beecher2025reviewerfatigue, mumcu2026agentic}. In response, journals, conferences, and funding agencies have begun to adopt author-level submission quotas \cite{AnnualAuthor}. A basic fixed quota assigns the same cost to every author of every submission, regardless of whether the paper is solo-authored or collaborative. This is simple, but it ignores coauthorship structure. A senior researcher involved in many student-led projects, for example, may exhaust their quota and thereby block junior collaborators from submitting otherwise independent work.

Recent coauthorship-sensitive quota rules address this issue by reducing the quota cost assigned to each author as the number of coauthors increases. A natural per-capita rule charges each author $1/a$ for a paper with $a$ authors, but this can be manipulated by adding spurious coauthors. Recent work has proposed harmonic and generalized harmonic quota rules to account for coauthorship while limiting author-list padding~\cite{shah2026many}. Harmonic quota rules offer a middle ground: each coauthor pays a cost that decreases with the harmonic number of the author count. This makes collaborative papers cheaper per author than solo papers while limiting the benefit of artificially expanding the author list.

This paper argues that such rules, while mathematically well motivated, rely on an incomplete model of academic submissions. In particular, they use authorship-credit dilution as a proxy for quota cost. This creates a mismatch between the object being modeled and the policy goal. Authorship credit, review burden, and submission responsibility are related but distinct. Review burden is primarily imposed by the submitted paper. Authorship credit concerns academic benefit. Submission responsibility concerns who is driving, preparing, and choosing to submit the work. A quota rule that uses only the number of coauthors cannot distinguish these dimensions, just as author lists alone are often insufficient to represent individual scholarly contributions~\cite{tscharntke2007authorsequence,brand2015beyondauthorship,credittaxonomy}.

This distinction matters in several common cases. First, genuine solo-authored papers receive the highest quota cost, even when the author is not gaming the system and is simply conducting independent work. Second, all coauthors of a paper are charged symmetrically, even though the first author, corresponding author, senior advisor, and secondary contributor may have very different roles. Third, in student-led research, a faculty advisor may be a necessary intellectual and supervisory contributor, but not the main source of submission volume. A coauthor-symmetric quota can therefore make the advisor's unrelated submissions a bottleneck for multiple students.

We propose that submission quotas should be role-aware. Instead of assigning the same quota cost to every coauthor solely as a function of author count, a quota mechanism should assign author-specific costs based on declared and constrained roles. Lead or submitting authors may bear the main responsibility cost; ordinary coauthors may bear a smaller collaboration cost; and designated advisors on student-led papers may receive a reduced supervisory cost. Such a framework can still control venue-level submission load, but it better reflects how research papers are produced and submitted.

The goal is not to remove submission limits or to make quotas easy to evade. Rather, the goal is to align quota costs with the reasons quotas are introduced. If the objective is to prevent individuals from overwhelming a venue with submissions, the cost should fall primarily on those driving the submissions. If the objective is to account for academic credit, credit models may be useful, but they should not be treated as the sole basis for submission responsibility. If the objective is to protect reviewing capacity, then the mechanism should control total submission load without unnecessarily penalizing solo authors or student-led work.

This paper makes the following contributions:
\begin{itemize}
    \item We identify modeling limitations of coauthor-symmetric quota rules, with emphasis on the distinction between review burden, authorship credit, and submission responsibility.

    \item We show that quota rules based only on author count can structurally disadvantage solo authors and student-led advisor-supervised papers.

    \item We formulate desiderata for submission quota mechanisms, including venue-load control, padding resistance, role sensitivity, solo neutrality, and student non-blocking.

    \item We propose a role-aware quota framework that assigns author-specific quota costs according to constrained roles such as lead author, regular coauthor, and designated advisor.

    \item We discuss how role constraints can preserve resistance to manipulation while avoiding several limitations of purely coauthor-symmetric quota rules.
\end{itemize}

The rest of the paper is organized as follows. Section~\ref{sec:background} reviews fixed, per-capita, harmonic, and generalized harmonic quota rules. Section~\ref{sec:limitations} analyzes their limitations. Section~\ref{sec:desiderata} introduces desiderata for role-aware submission quotas. Section~\ref{sec:framework} presents the proposed framework. Section~\ref{sec:properties} discusses basic properties and manipulation resistance. Section~\ref{sec:scenarios} compares quota behavior across representative authorship scenarios. Section~\ref{sec:conclusion} concludes.

\section{Background: Coauthor-Symmetric Quota Rules}
\label{sec:background}

We first review several coauthor-symmetric quota rules, where the cost of a submission is assigned equally to all coauthors, following the fixed, per-capita, harmonic, and generalized harmonic formulations introduced in~\cite{shah2026many}. Let a submitted paper have $a \geq 1$ authors. A coauthor-symmetric quota rule assigns the same quota cost $f(a)$ to each author of that paper, where $f$ depends only on the number of authors. Each author has an annual budget $B$, and a submission is permitted only if every coauthor has at least $f(a)$ remaining budget.

\subsection{Fixed Quota Rule}

The fixed quota rule assigns unit cost to every author of every submission:
\begin{equation}
    f_{\mathrm{fixed}}(a) = 1.
\end{equation}
Under this rule, each author may appear on at most $B$ submissions, regardless of whether the papers are solo-authored or collaborative. The rule is simple and transparent, but it ignores coauthorship structure. In particular, a solo-authored paper and a large collaborative paper consume the same quota from each listed author.

\subsection{Per-Capita Quota Rule}

A natural alternative is to divide the submission cost equally among all coauthors:
\begin{equation}
    f_{\mathrm{capita}}(a) = \frac{1}{a}.
\end{equation}
This rule treats a submission as a unit cost distributed across its authors. It recognizes that multi-author papers are collaborative and therefore charges each author less than in the fixed quota rule. However, because the cost decreases linearly with the number of authors, the rule is vulnerable to author-list padding: adding extra coauthors can substantially reduce each author's quota cost.

\subsection{Harmonic Quota Rule}

The harmonic quota rule was proposed to interpolate between the fixed and per-capita approaches. Let
\begin{equation}
    H_a = \sum_{j=1}^{a} \frac{1}{j}
\end{equation}
denote the $a$-th harmonic number. In its basic form, the harmonic quota cost decreases as
\begin{equation}
    f_{\mathrm{harm}}(a) = \frac{1}{H_a}.
\end{equation}
Since $H_a$ grows logarithmically, the cost decreases more slowly than $1/a$. Thus, collaborative papers are cheaper per author than solo papers, but adding authors provides diminishing quota benefit.

A more general version introduces an irreducible per-author component. Given parameters $N_1$ and $N_{\infty}$, where $N_1$ is the number of allowed single-author submissions and $N_{\infty}$ is the limiting number of submissions for an author appearing only on very large collaborations, the cost is
\begin{equation}
    f_{\mathrm{harm}}(a)
    =
    \frac{N_1}{N_{\infty}}
    +
    \left(1 - \frac{N_1}{N_{\infty}}\right)\frac{1}{H_a}.
\end{equation}
Each author begins with budget $N_1$. A submission is allowed if every coauthor has at least $f_{\mathrm{harm}}(a)$ remaining budget, after which this amount is deducted from every coauthor.

\subsection{Generalized Harmonic Quota Rule}

The generalized harmonic quota rule further extends this framework by allowing organizers to specify a finite collaboration size $A$ and the number of allowed submissions $N_A$ for authors whose papers all have $A$ authors. This yields a family of rules based on generalized harmonic numbers
\begin{equation}
    H_{a,p} = \sum_{j=1}^{a} \frac{1}{j^p},
\end{equation}
where the parameter $p$ controls how quickly quota cost decreases with author count. The resulting cost has the form
\begin{equation}
    f_{\mathrm{gen}}(a)
    =
    \frac{N_1}{N_{\infty}}
    +
    \left(1 - \frac{N_1}{N_{\infty}}\right)\frac{1}{H_{a,p}},
\end{equation}
with $p$ and $N_{\infty}$ chosen to satisfy the organizer-specified parameters. This framework includes fixed, per-capita, and harmonic quota rules as special or limiting cases.

\subsection{Common Structure}

All of the above rules are coauthor-symmetric: every listed author of a paper pays the same quota cost. They differ only in how that common cost depends on the number of authors. This structure makes the rules simple and easy to compute, but it also means that author roles are not represented. A first author, corresponding author, faculty advisor, student lead, and secondary contributor are treated identically for quota purposes whenever they appear on the same paper.

The next section analyzes the limitations that arise from this symmetry.

\section{Limitations of Coauthor-Symmetric Harmonic Quotas}
\label{sec:limitations}

Coauthor-sensitive quota rules improve over fixed quotas by recognizing that collaboration structure matters. However, rules based only on the number of coauthors remain limited. The central issue is that harmonic quota rules use authorship-credit dilution as the basis for quota cost, while submission quotas are primarily intended to regulate review load and submission behavior.

\textbf{Student-led papers can be blocked by advisor quota.}
In Ph.D. research, many papers are primarily student-led but include a faculty advisor as a legitimate supervisory and intellectual contributor. If every coauthor is charged symmetrically, the advisor's quota can become a bottleneck across unrelated student projects. This can block students who have remaining quota simply because their advisor appears on several other student-led submissions.

\textbf{Credit is not review burden.}
Harmonic quota rules are motivated by models of authorship credit, where the credit of an author decreases as the number of coauthors increases. However, review burden is mostly imposed by the submitted paper itself. A paper generally requires reviewers, editorial handling, discussion, and a decision regardless of whether it has one author or many authors. Therefore, authorship-credit dilution is not a sufficient proxy for the reviewing cost imposed on the venue.

\textbf{Credit is not submission responsibility.}
Submission responsibility is also distinct from authorship credit. In many papers, the first author, corresponding author, or submitting author is primarily responsible for preparing and submitting the work. Other coauthors may contribute experiments, data, analysis, advising, or feedback. Coauthor-symmetric rules charge all of these roles equally, even though they do not represent equal responsibility for the act of submission.

\textbf{Solo-authored work is structurally penalized.}
Under harmonic-style rules, solo-authored papers receive the maximum quota cost, while multi-author papers receive a lower per-author cost. This may be consistent with full authorship credit, but it can disadvantage genuine independent work. A researcher who writes a paper alone is not necessarily abusing the system or imposing more reviewing burden than a researcher submitting collaborative papers.

\textbf{The anti-padding guarantee is narrow.}
The harmonic rule provides a formal argument against one manipulation mode: adding spurious coauthors to reduce quota cost. This is useful, but it does not establish broader fairness. It does not address role differences, student-led submissions, solo-author neutrality, or the distinction between review burden and authorship credit.

\textbf{Author-list expansion can still increase submission capacity.}
Even when harmonic quotas reduce the benefit of adding coauthors compared with per-capita counting, the per-author quota cost still decreases as the author count grows. Therefore, a lead or submitting author can obtain greater submission capacity by appearing on papers with more coauthors. This does not necessarily imply spurious authorship, but it shows that the mechanism still rewards larger author lists at the quota level. A quota rule should not create a direct submission-capacity advantage from adding coauthors when the same author remains responsible for driving the submission.

Overall, the limitation is not merely the harmonic function itself, but the assumption that author count alone is sufficient to determine quota cost. A more suitable mechanism should separate review burden, authorship credit, and submission responsibility, and should assign quota costs in a role-aware manner.

\section{Desiderata for Submission Quota Rules}
\label{sec:desiderata}

We now state desiderata for submission quota rules. These properties are not intended to define a unique mechanism. Rather, they clarify what a quota rule should preserve when limiting submission volume.

\textbf{Venue-load control.}
A quota rule should bound the total number of submissions an individual can drive within a given period. This is the primary reason for introducing submission quotas: to reduce excessive reviewing and editorial load.

\textbf{Padding resistance.}
Adding spurious coauthors should not increase an author's effective submission capacity. If quota cost decreases with the number of authors, the decrease must be constrained so that authors do not benefit from artificially expanding the author list.

\textbf{Role sensitivity.}
Quota cost should reflect author roles. A lead, corresponding, or submitting author generally has greater responsibility for a submission than a peripheral contributor or a supervisory advisor. A rule that charges all coauthors equally cannot represent this distinction.

\textbf{Solo neutrality.}
A quota rule should not unnecessarily disadvantage genuine solo-authored work. Solo papers may assign full academic credit to one author, but this does not imply that solo authors should be more restricted than lead authors of collaborative papers with comparable submission responsibility.

\textbf{Student non-blocking.}
In student-led work, a student's ability to submit should not be primarily determined by the advisor's unrelated submissions. A quota rule should allow venues to distinguish the student or lead author who drives the submission from a faculty advisor who provides supervisory and intellectual contribution.

\textbf{Transparency.}
Authors should be able to compute their quota usage before submission. A quota mechanism that is too complex or depends on unverifiable hidden quantities may be difficult to enforce consistently.

\textbf{Manipulation awareness.}
Role-aware mechanisms should anticipate strategic behavior, such as mislabeling author roles, rotating corresponding authors, or adding nominal advisors. Therefore, role declarations should be constrained, auditable, and simple enough for venues to enforce.

Together, these desiderata suggest that a quota mechanism should not rely only on author count. It should control submission volume while distinguishing the different roles through which authors contribute to, supervise, and drive submissions.

\section{Role-Aware Quota Framework}
\label{sec:framework}

We propose a role-aware quota framework in which quota cost is assigned at the author level rather than only at the paper level. Let $P$ be a submitted paper with author set $\mathcal{A}(P)$. Instead of assigning the same cost $f(|\mathcal{A}(P)|)$ to every coauthor, the rule assigns each author $i \in \mathcal{A}(P)$ an individual quota cost based on a declared quota role.

Let
\begin{equation}
    z_i^{\mathrm{lead}}(P),
    z_i^{\mathrm{co}}(P),
    z_i^{\mathrm{adv}}(P)
    \in \{0,1\}
\end{equation}
denote one-hot-encoded binary indicators specifying whether author $i$ is a lead or submitting author, regular coauthor, or designated advisor on paper $P$. The role-aware quota cost is
\begin{equation}
    q_i(P)
    =
    q_{\mathrm{lead}}(P) z_i^{\mathrm{lead}}(P)
    +
    q_{\mathrm{co}}(P) z_i^{\mathrm{co}}(P)
    +
    q_{\mathrm{adv}}(P) z_i^{\mathrm{adv}}(P).
    \label{eq:role-aware-cost}
\end{equation}

The role costs may be fixed constants or venue-defined functions of paper attributes such as the number of authors. For example, setting all role costs equal to the same function $f(|\mathcal{A}(P)|)$ recovers a coauthor-symmetric quota rule.

In the default role-exclusive version, each author is assigned exactly one quota role:
\begin{equation}
    z_i^{\mathrm{lead}}(P)
    +
    z_i^{\mathrm{co}}(P)
    +
    z_i^{\mathrm{adv}}(P)
    =
    1
    \quad
    \text{for all } i \in \mathcal{A}(P).
\end{equation}
This avoids an unconstrained weighted sum of overlapping responsibilities and keeps quota usage transparent. The role-aware view is consistent with contributor-taxonomy approaches such as CRediT, which represent scholarly work through distinct contribution categories including supervision, writing, investigation, methodology, and project administration~\cite{brand2015beyondauthorship,credittaxonomy,elseviercredit}.

\textbf{Lead or submitting author.}
The lead or submitting author role captures primary responsibility for driving the submission. This may correspond to the first author, corresponding author, submitting author, or another explicitly declared lead. Since submission quotas are intended to limit excessive submissions, this role receives the largest quota cost.

\textbf{Regular coauthor.}
The regular coauthor role captures ordinary coauthorship contribution without primary submission responsibility. Such contributors may provide experiments, data, analysis, writing, feedback, or other intellectual contributions. Unlike coauthor-symmetric harmonic quotas, this role does not determine quota cost solely through author count; instead, it assigns a reduced cost to authors who contribute to the paper but are not responsible for driving the submission.

\textbf{Designated advisor.}
The designated advisor role captures faculty advising or supervision, especially in student-led work. A designated advisor may receive a smaller quota cost than the student or lead author, reflecting the advisor's legitimate supervisory and intellectual contribution without making the advisor the primary bottleneck for unrelated student submissions.

A submission is permitted if every author has sufficient remaining budget for their assigned role cost:
\begin{equation}
    B_i \geq q_i(P)
    \quad
    \text{for all } i \in \mathcal{A}(P),
\end{equation}
where $B_i$ is author $i$'s remaining quota budget. After submission, each author's budget is updated as
\begin{equation}
    B_i \leftarrow B_i - q_i(P).
\end{equation}
Importantly, the framework does not require separate budgets for different author roles. Each author may have a single quota budget, while different role assignments deduct different amounts from that same budget.

\textbf{Role costs.}
Since the leading author is supposed to spend significantly more time than other co-authors, it is natural to expect less submissions from someone who typically leads a paper (e.g., a PhD student) compared to an advisor. Hence, we propose the following ranking for the role costs:
\begin{equation}
    q_{\mathrm{lead}} \geq q_{\mathrm{co}} \geq q_{\mathrm{adv}} \geq 0.
\end{equation}

\textbf{Role constraints.}
To limit manipulation, role declarations should be constrained by simple and easy-to-verify rules. For example, a paper may have at most one designated advisor, at least one lead or submitting author, and a fixed set of allowed role labels. Advisor status should require an externally verifiable senior or supervisory position, such as faculty, research scientist, principal investigator, group leader, or postdoctoral mentor status, depending on the venue's policy. In particular, a student author should not be eligible for the advisor role merely to reduce their quota cost.

\section{Properties and Special Cases}
\label{sec:properties}

We now discuss several basic properties of the role-aware framework. These properties follow directly from the definition of role-specific quota costs and are intended to clarify how the framework relates to existing coauthor-symmetric rules.

\textbf{Existing quota rules as special cases.}
The role-aware framework contains fixed, per-capita, harmonic, and generalized harmonic quota rules as special cases. For a submitted paper $P$ with $a = |\mathcal{A}(P)|$ authors, this is obtained by assigning the same quota cost to every role. In other words, the mechanism ignores role differences by setting
\begin{equation}
    q_{\mathrm{lead}}(a)
    =
    q_{\mathrm{co}}(a)
    =
    q_{\mathrm{adv}}(a)
    =
    f(a).
\end{equation}
If $f(a)=1$, then every author pays unit cost and the fixed quota rule is recovered. If $f(a)=1/a$, then every author pays an equal share of the submission cost and the per-capita quota rule is recovered. If $f(a)=1/H_a$, then every author pays the harmonic quota cost. Similarly, setting
\begin{equation}
    f(a)=\frac{1}{H_{a,p}}
\end{equation}
recovers the generalized harmonic form used in our scenario analysis, and the more general shifted harmonic form can be obtained by setting $f(a)$ equal to the corresponding cost function. Thus, coauthor-symmetric quota rules are recovered when all declared roles are assigned the same cost.

\textbf{Role sensitivity.}
The framework departs from coauthor-symmetric rules by allowing different roles to have different costs. A simple ordering is
\begin{equation}
    q_{\mathrm{lead}} \geq q_{\mathrm{co}} \geq q_{\mathrm{adv}} \geq 0.
\end{equation}
This makes the quota cost monotone with declared submission responsibility: lead or submitting authors pay the largest cost, regular coauthors pay an intermediate cost, and designated advisors may pay a reduced supervisory cost. The ordering is a design choice rather than a mathematical necessity, but it directly reflects the intended distinction between driving a submission, contributing to it, and supervising it.

\textbf{Lead-author padding resistance.}
In the default role-aware rule, the lead or submitting author pays a fixed responsibility cost $q_{\mathrm{lead}}$ that does not decrease when additional coauthors are added. Therefore, adding coauthors cannot increase the lead author's submission capacity by reducing the lead-author cost. This addresses a limitation of coauthor-symmetric rules, where the quota cost of every author decreases as the author count grows.

\textbf{Solo neutrality.}
The framework can avoid penalizing solo-authored work by assigning solo papers the same responsibility cost as lead-authored collaborative papers:
\begin{equation}
    q_{\mathrm{solo}} = q_{\mathrm{lead}}.
\end{equation}
Under this choice, a solo author and the lead author of a collaborative paper consume the same quota cost. This separates submission responsibility from collaboration size and prevents solo authorship from being automatically treated as the most costly case.

\textbf{Student non-blocking.}
For student-led papers, the framework can reduce advisor bottlenecks by assigning the student lead the main responsibility cost and the advisor a smaller supervisory cost:
\begin{equation}
    q_{\mathrm{student}} = q_{\mathrm{lead}},
    \qquad
    q_{\mathrm{advisor}} = \lambda q_{\mathrm{lead}},
    \qquad
    0 \leq \lambda < 1.
\end{equation}
This does not remove quota constraints. The student lead still pays the main submission cost. However, it prevents the advisor's unrelated submissions from automatically blocking multiple independent student-led papers.

\textbf{Shared lead responsibility.}
The optional shared-lead extension allows a small number of authors to share the lead responsibility cost. This can represent cases where two authors genuinely share submission responsibility. However, unrestricted sharing would be vulnerable to manipulation, since authors could reduce the lead cost by declaring many co-leads. Therefore, if shared lead responsibility is allowed, the number of lead authors should be capped, for example at two. Under such a cap, shared responsibility remains possible without making the lead-author cost arbitrarily small:
\begin{equation}
    q_1=q_2=\frac{q_{lead}}{2}+\frac{q_{co}}{2}.
\end{equation}

Overall, these properties show that role-aware quotas preserve the standard quota rules as limiting cases while allowing venues to assign costs according to declared responsibility. The framework can behave exactly like a coauthor-symmetric rule when all role costs are equal, or it can distinguish lead authors, regular coauthors, and advisors when role sensitivity is desired.

\section{Scenario Analysis}
\label{sec:scenarios}

\begin{figure}[t]
    \centering

    \includegraphics[width=0.65\linewidth]{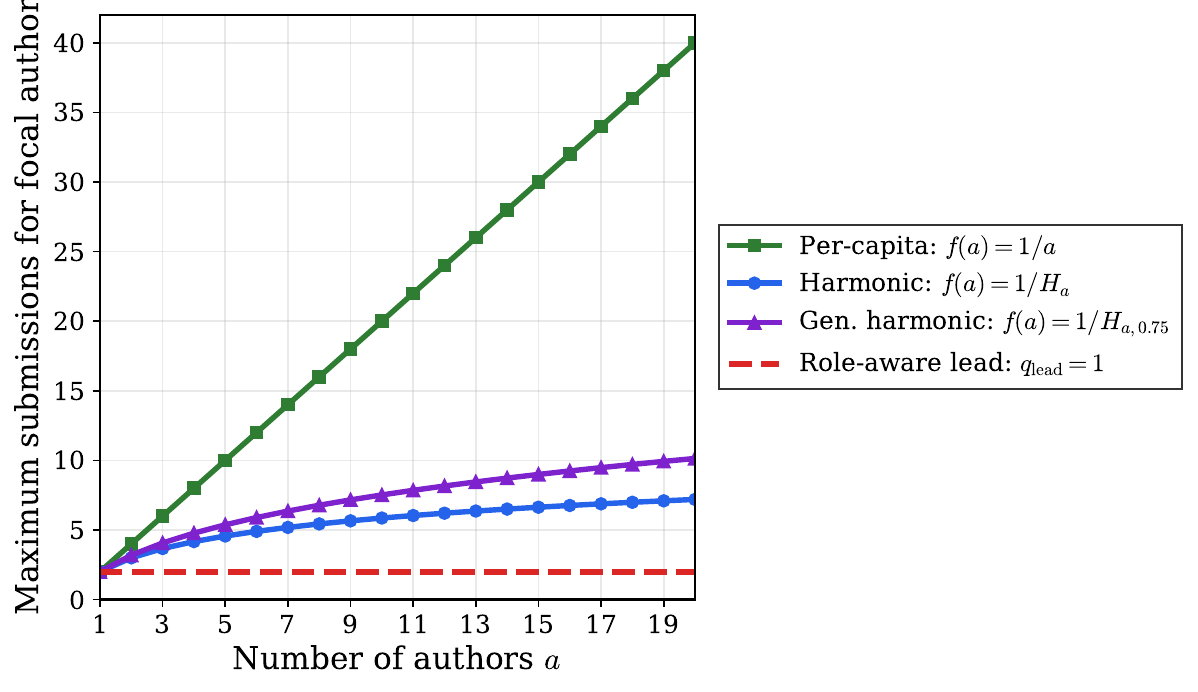}
    \caption{Maximum submission capacity as a function of author count.}
    \label{fig:capacity-author-count}

    \vspace{0.8em}

    \includegraphics[width=0.65\linewidth]{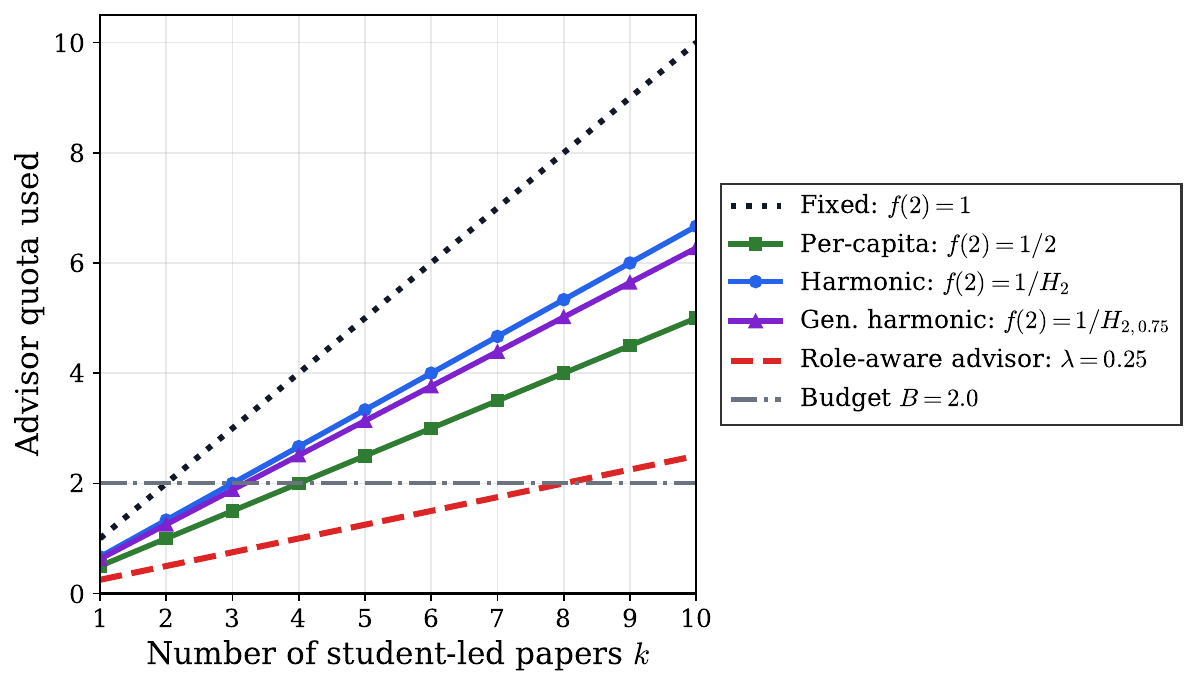}
    \caption{Cumulative advisor quota usage across student-led papers.}
    \label{fig:advisor-quota-usage}
\end{figure}

We compare quota behavior under deterministic synthetic authorship scenarios. The goal is not to model a particular venue or estimate real submission volume, but to isolate structural differences between quota mechanisms. We use an annual budget $B=2$ and compare fixed, per-capita, harmonic, generalized harmonic, and role-aware quota rules where applicable. For the role-aware rule, we set $q_{\mathrm{lead}}=1$ as a normalization and use reduced costs for non-lead roles.

Figure~\ref{fig:capacity-author-count} shows the maximum submission capacity of a focal author as the number of authors increases. For coauthor-symmetric rules, the focal author pays a cost that decreases with author count. As a result, per-capita, harmonic, and generalized harmonic rules all increase the focal author's submission capacity as more coauthors are added. The effect is strongest under per-capita counting and weaker under harmonic counting, but it is not eliminated. In contrast, the role-aware lead-author rule keeps the lead or submitting author's cost fixed. Thus, adding coauthors does not increase the submission capacity of the author driving the paper.

Figure~\ref{fig:advisor-quota-usage} considers student-led papers that share the same faculty advisor. Each paper has a different student lead and the same advisor. Under coauthor-symmetric rules, the advisor is charged on every paper in the same way as any other coauthor. Consequently, the advisor's cumulative quota usage grows quickly and can become a bottleneck across unrelated student-led projects. Under the role-aware rule, the student lead pays the main responsibility cost, while the advisor pays a reduced supervisory cost. This weakens the advisor bottleneck without removing the quota constraint on the student lead.

\begin{figure}[t]
    \centering
    \includegraphics[width=0.65\linewidth]{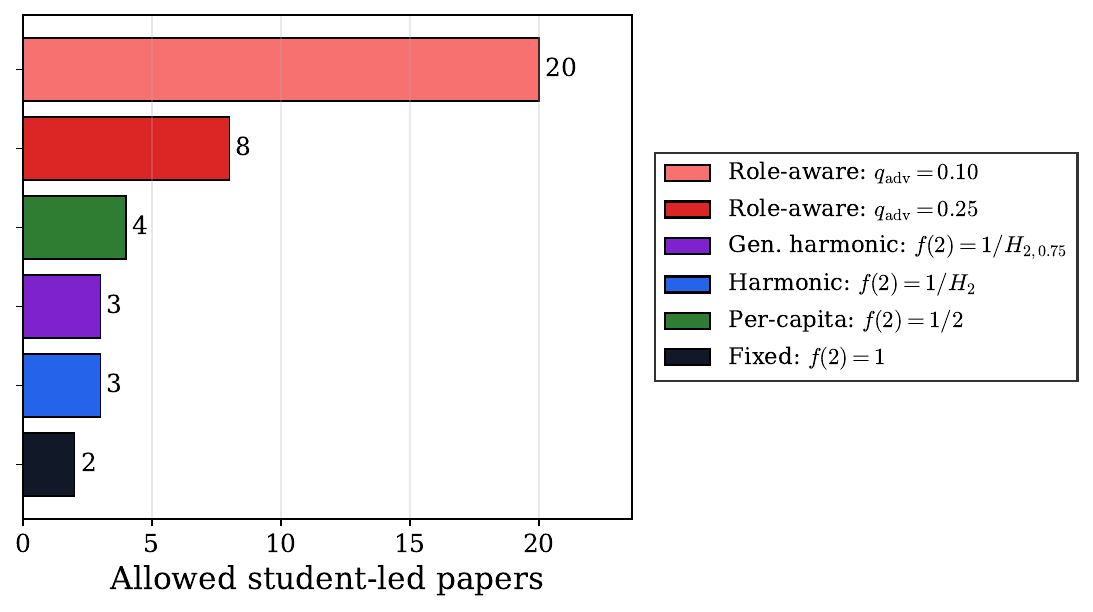}
    \caption{Allowed student-led papers before advisor quota becomes binding.}
    \label{fig:advisor-blocking}
\end{figure}

Figure~\ref{fig:advisor-blocking} reports the number of independent student-led papers that can be submitted before the shared advisor's quota blocks the next submission. This is the discrete consequence of the cumulative behavior in Figure~\ref{fig:advisor-quota-usage}. Fixed and coauthor-symmetric rules allow only a small number of student-led papers before the advisor becomes the limiting constraint. Role-aware rules allow more student-led submissions from the same advising group, but the additional capacity comes from reducing the advisor's supervisory cost, not from removing the lead author's quota. Each student lead remains subject to the main submission-responsibility cost.

Together, these scenarios show two structural effects. First, coauthor-symmetric rules can make submission capacity depend on author count, creating a quota-level advantage for larger author lists. Second, the same symmetry can make a faculty advisor a bottleneck for independent student-led papers. A role-aware rule addresses both issues by assigning the main quota cost to the author driving the submission while allowing smaller role-specific costs for coauthors and advisors.

\section{Conclusion}
\label{sec:conclusion}

Author-level submission quotas are a practical response to increasing peer-review load, but their design requires care. Coauthor-sensitive rules such as per-capita and harmonic quotas improve over fixed per-author limits by recognizing that collaborative papers should not always consume the same quota from every author. However, we argued that coauthor-symmetric rules remain limited because they determine quota cost only from the number of authors. This conflates authorship credit, review burden, and submission responsibility.

We identified several structural limitations of author-count-based quota rules. Such rules can make an author's effective submission capacity increase with the number of coauthors, even when the same author remains responsible for driving the submission. They can also penalize genuine solo-authored work and create advisor bottlenecks in student-led research, where multiple independent student projects share the same faculty advisor. Although harmonic quota rules reduce the incentive for simple author-list padding, they do not address these broader role-based differences.

To address these issues, we proposed a role-aware quota framework. The framework assigns author-specific quota costs based on constrained roles such as lead author, regular coauthor, and designated advisor. Existing fixed, per-capita, and harmonic rules appear as special cases, while the more general formulation allows venues to distinguish lead authors, regular coauthors, and advisors. Our deterministic scenario analysis shows that role-aware costs can prevent submission capacity from increasing merely through additional coauthors and can reduce advisor bottlenecks for independent student-led papers.

The goal of role-aware quotas is not to eliminate submission limits or to favor large research groups. Rather, it is to align quota costs with the reason quotas are introduced: limiting excessive submissions by the individuals responsible for driving them. By separating submission responsibility from authorship credit and supervision, role-aware mechanisms provide a more flexible and faithful foundation for managing submission volume under modern collaborative authorship.

\bibliographystyle{plainnat}
\bibliography{references}

\end{document}